\documentclass[aps,prl,twocolumn,superscriptaddress]{revtex4-1}
\usepackage{graphicx}
\usepackage[hidelinks]{hyperref} 
\usepackage{amsmath}
\usepackage{bm}
\usepackage{color}
\usepackage{lineno}

\usepackage{etoolbox}

\usepackage{lipsum}
\graphicspath{{./}}

\begin{document}


\title{Fractional antiferromagnetic skyrmion lattice induced by anisotropic couplings}

\author{Shang Gao}
\thanks{Present address: Materials Science \& Technology Division and Neutron Science Division, Oak Ridge National Laboratory, Oak Ridge, TN 37831, USA}
\affiliation{Laboratory for Neutron Scattering and Imaging, Paul Scherrer Institut, CH-5232 Villigen PSI, Switzerland}
\affiliation{Department of Quantum Matter Physics, University of Geneva, CH-1211 Geneva, Switzerland}
\affiliation{RIKEN Center for Emergent Matter Science, Wako 351-0198, Japan}

\author{H. D. Rosales}
\affiliation{Instituto de F\'isica de L\'iquidos y Sistemas Biol\'ogicos, CCT La Plata, CONICET and Departamento de F\'isica, Facultad de Ciencias Exactas, Universidad Nacional de La Plata, C.C. 67, 1900 La Plata, Argentina}
\affiliation{Departamento de Ciencias Básicas, Facultad de Ingeniería, Universidad Nacional de La Plata, C.C. 67, 1900 La Plata, Argentina}

\author{F. A. G\'omez Albarrac\'in}
\affiliation{Instituto de F\'isica de L\'iquidos y Sistemas Biol\'ogicos, CCT La Plata, CONICET and Departamento de F\'isica, Facultad de Ciencias Exactas, Universidad Nacional de La Plata, C.C. 67, 1900 La Plata, Argentina}
\affiliation{Departamento de Ciencias Básicas, Facultad de Ingeniería, Universidad Nacional de La Plata, C.C. 67, 1900 La Plata, Argentina}

\author{Vladimir Tsurkan}
\affiliation{Experimental Physics V, University of Augsburg, D-86135 Augsburg, Germany}
\affiliation{Institute of Applied Physics, MD-2028 Chisinau, Republic of Moldova}

\author{Guratinder Kaur}
\affiliation{Laboratory for Neutron Scattering and Imaging, Paul Scherrer Institut, CH-5232 Villigen PSI, Switzerland}
\affiliation{Department of Quantum Matter Physics, University of Geneva, CH-1211 Geneva, Switzerland}

\author{Tom Fennell}
\affiliation{Laboratory for Neutron Scattering and Imaging, Paul Scherrer Institut, CH-5232 Villigen PSI, Switzerland}

\author{Paul Steffens}
\affiliation{Institut Laue-Langevin, CS 20156, 38042 Grenoble Cedex 9, France}

\author{Martin Boehm}
\affiliation{Institut Laue-Langevin, CS 20156, 38042 Grenoble Cedex 9, France}

\author{Petr \v{C}erm\'{a}k}
\affiliation{J\"ulich Center for Neutron Science at Heinz Maier-Leibnitz Zentrum, Forshungszentrum J\"ulich GmbH,  D-85747 Garching, Germany}
\affiliation{Department of Condensed Matter Physics, Faculty of Mathematics and Physics, Charles University, Ke Karlovu 5, 121 16, Praha, Czech Republic}

\author{Astrid Schneidewind}
\affiliation{J\"ulich Center for Neutron Science at Heinz Maier-Leibnitz Zentrum, Forshungszentrum J\"ulich GmbH,  D-85747 Garching, Germany}

\author{Eric Ressouche}
\affiliation{Universit\'e Grenoble Alpes, CEA, INAC-MEM, F-38000 Grenoble, France}

\author{Daniel C. Cabra}
\affiliation{Instituto de F\'isica de L\'iquidos y Sistemas Biol\'ogicos, CCT La Plata, CONICET and Departamento de F\'isica, Facultad de Ciencias Exactas, Universidad Nacional de La Plata, C.C. 67, 1900 La Plata, Argentina}
\affiliation{Abdus Salam International Centre for Theoretical Physics, Associate Scheme, Strada Costiera 11, I-34151, Trieste, Italy}

\author{Christian R\"uegg}
\affiliation{Department of Quantum Matter Physics, University of Geneva, CH-1211 Geneva, Switzerland}
\affiliation{Neutrons and Muons Research Division, Paul Scherrer Institut, CH-1211 Villigen PSI, Switzerland}
\affiliation{Institute for Quantum Electronics, ETH Z\"urich, CH-8093 Z\"urich, Switzerland}
\affiliation{Institute of Physics, \'Ecole Polytechnique F\'ed\'erale de Lausanne, CH-1015 Lausanne, Switzerland}

\author{Oksana Zaharko}
\email[]{oksana.zaharko@psi.ch}
\affiliation{Laboratory for Neutron Scattering and Imaging, Paul Scherrer Institut, CH-5232 Villigen PSI, Switzerland}


\date{\today}


\begin{abstract}
Magnetic skyrmions are topological solitons with a nanoscale winding spin texture that hold promise for spintronics applications~\cite{muhlbauer_skyrmion_2009, yu_real_2010, nagaosa_topological_2013, fert_magnetic_2017}. Until now, skyrmions have been observed in a variety of magnets that exhibit nearly parallel alignment for the neighbouring spins, but theoretically, skyrmions with anti-parallel neighbouring spins are also possible. The latter, antiferromagnetic skyrmions, may allow more flexible control compared to the conventional ferromagnetic skyrmions~\cite{baltz_anti_2018, barker_static_2016, zhang_antiferromagnetic_2016, rosales_three_2015, diaz_topological_2019, kamiya_magnetic_2014}. Here, by combining neutron scattering and Monte Carlo  simulations, we show that a fractional antiferromagnetic skyrmion lattice with an incipient meron character~\cite{lin_skyrmion_2015, yu_transformation_2018} is stabilized in MnSc$_2$S$_4$ through anisotropic couplings. Our work demonstrates that the theoretically proposed antiferromagnetic skyrmions can be stabilized in real materials and represents an important step towards implementing the antiferromagnetic-skyrmion based spintronic devices.
\end{abstract}

\maketitle

The concept of topology has revolutionized condensed matter physics: it reveals that the classification of different phases can extend beyond the Landau-Ginzburg-Wilson paradigm of classification by symmetry, bringing about a variety of new phases with topological characters~\cite{wen_zoo_2017}. Among the topological entities, magnetic skyrmions with a winding spin texture in real space have triggered enormous interest due to their potential for spintronics applications~\cite{muhlbauer_skyrmion_2009, yu_real_2010, nagaosa_topological_2013, fert_magnetic_2017}. Information encoded in the nanoscale spin winding of the skyrmions is topologically protected against perturbations, and can be conveniently manipulated with electronic currents~\cite{jonietz_spin_2010, yu_skyrmion_2012, white_electric_2012}.

Similar to the vortices that emerge in the Berezinskii-Kosterlitz-Thouless transition, magnetic skymions are conventionally treated as topological solitons in non-linear field theory~\cite{roszler_spontaneous_2006}, which implies a continuous ferromagnetic spin alignment at short length scales. This short-range ferromagnetism is indeed a common feature for most of the known skyrmion hosts, including the chiral magnets with antisymmetric Dzyaloshinskii-Moriya interactions (DMI)~\cite{nagaosa_topological_2013}, and the recently discovered centrosymmetric compounds with multiple-spin couplings~\cite{kurumaji_skyrmion_2019,hirschberger_skyrmion_2018, khanh_nanometric_2020}. 
\begin{figure}[t!]
\includegraphics[width=0.48\textwidth]{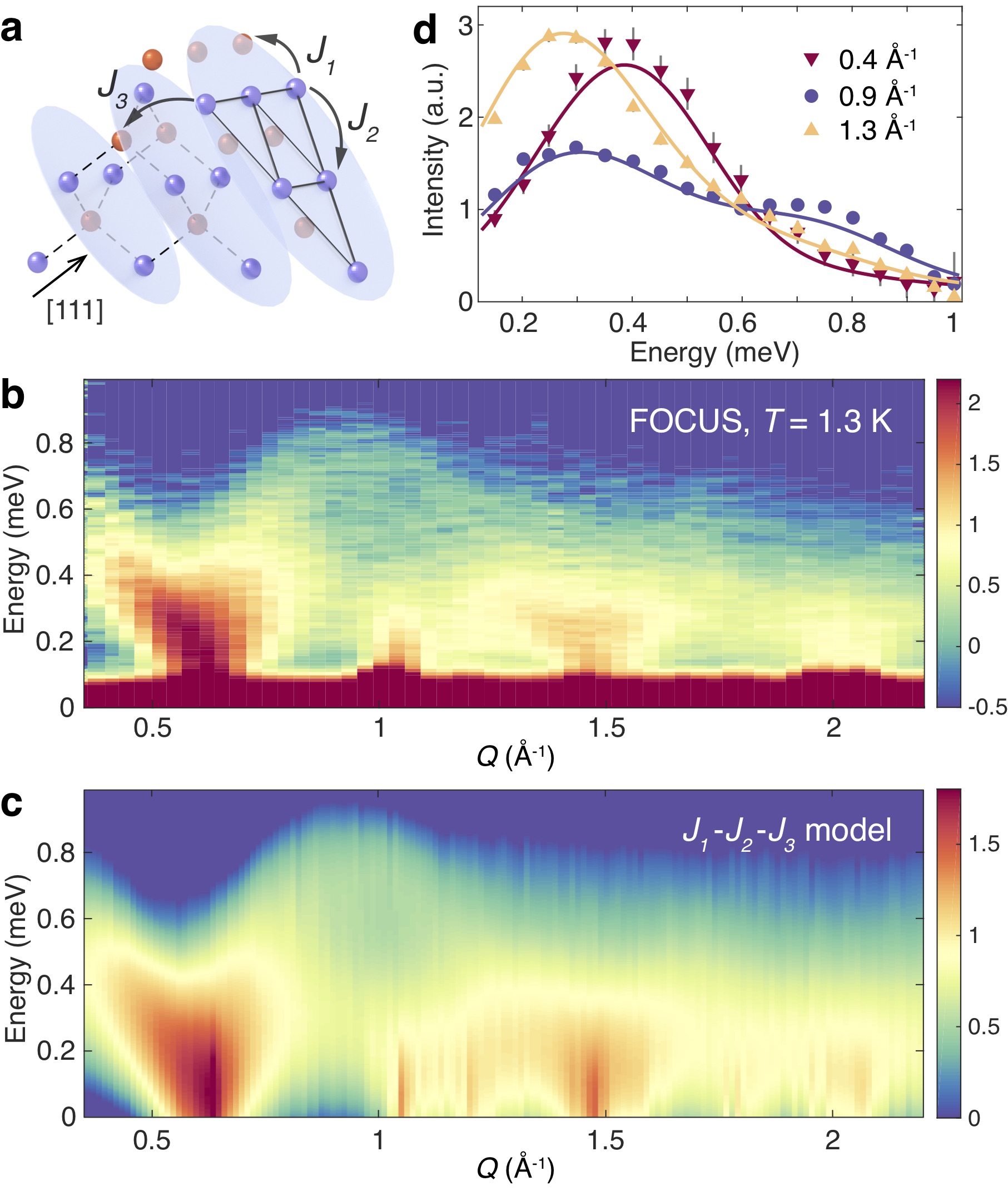}
\caption{\textbf{Spin dynamics in a powder sample of MnSc$_2$S$_4$. a,} Mn$^{2+}$ ions (blue and brown spheres) in MnSc$_2$S$_4$ form a bipartite diamond lattice that can be viewed as triangular planes (blue) stacked along the [111] direction. Couplings up to the third neighbours are indicated. The presentation of the magnetic lattice in the cubic unit cell can be found in Ref.~\citenum{gao_spiral_2017}. \textbf{b,} INS spectra $S(Q,\omega)$ collected on FOCUS at $T=$ 1.3 K using a powder sample of MnSc$_2$S$_4$. \textbf{c,} INS spectra calculated using the linear spin wave theory for the $J_1$-$J_2$-$J_3$ model with $J_1 = -0.31(1)$ K, $J_2=0.46(1)$ K, and $J_3=0.087(4)$ K. The calculated spectra are convoluted by a Gaussian function with fitted full-width-half-maximum (FWHM) of 0.27 meV to account for the instrumental resolution and thermal broadening. $\textbf{d,}$ Integrated INS spectra $I(\omega)$ at $Q=0.4$ (red triangles), $0.9$ (blue circles), and $1.3$ \AA$^{-1}$ (yellow triangles) with an integration width of 0.1 \AA$^{-1}$. Solid lines are the fitted spectra using the $J_1$-$J_2$-$J_3$ model at the corresponding $Q$ positions. Error bars represent standard deviations.
\label{fig:ins_powder}}
\end{figure}

However, explorations on skyrmions should not be confined to ferromagnets~\cite{sokolov_metamagnetic_2019}. Theoretical calculations have suggested that skyrmions might be also stabilized in antiferromagnets with two~\cite{barker_static_2016, zhang_antiferromagnetic_2016} or three~\cite{rosales_three_2015,diaz_topological_2019,kamiya_magnetic_2014} sublattices, leading to antiferromagnetic skyrmions (AF-Sks) with anti-parallel nearest-neighbouring (NN) spin alignment, which might complement the skyrmion control in spintronic devices~\cite{baltz_anti_2018}. On the other hand, antiferromagnets are often accompanied by strong frustration, which is a known ingredient to enhance fluctuations~\cite{balents_spin_2010}. Thus the marriage between skyrmion and antiferromagnetism~\cite{okubo_multiple_2012, leonov_multiply_2015} might be the key to realize exotic states like magnetic hopfions~\cite{sutcliffe_skyrmion_2017, rybakov_magnetic_2019} or even quantum skyrmions~\cite{lohani_quantum_2019}.

Despite their tantalizing prospects, it is as yet unclear whether AF-Sks can be experi-mentally realized or not. Direct observation of the AF-Sks, \textit{e.g.} with Lorentz transmission electron microscopy~\cite{yu_real_2010}, is challenging since the alternating spins cancel the local magnetic field. Although single-$\bm{q}$ magnetic structures can be accurately determined by neutron diffraction, skyrmion lattices are multi-$\bm{q}$ structures and the phase factors between the different propagation vectors $\bm{q}$ are lost.
One prominent example is the spinel MnSc$_2$S$_4$~\cite{fritsch_spin_2004,gao_spiral_2017}, where the magnetic Mn$^{2+}$ ions form a bipartite diamond lattice (see Fig.~\ref{fig:ins_powder}a). A previous neutron diffraction work revealed the existence of a field-induced triple-$\bm{q}$ phase in this antiferromagnet~\cite{gao_spiral_2017}, but the exact arrangement of magnetic moments still remains unclear.

In this article, we show that a fractional three-sublattice AF-Sk lattice is realized in the MnSc$_2$S$_4$ triple-$\bm{q}$ phase. By combining state-of-the-art neutron spectroscopy, extensive Monte Carlo simulations, and neutron diffraction, we clarify the microscopic couplings between the Mn$^{2+}$ spins in MnSc$_2$S$_4$ up to the third-neighbours and, most importantly, establish the existence of a fractional three-sublattice AF-Sk lattice~\cite{rosales_three_2015, diaz_topological_2019, kamiya_magnetic_2014} that originates from anisotropic couplings over the nearest-neighbours. The fractionalization of the AF-Sks can be attributed to their close packing~\cite{lin_skyrmion_2015}, leading to incomplete spin wrapping that is reminiscent of the magnetic merons/antimerons~\cite{yu_transformation_2018}.

Inelastic neutron scattering (INS) probes the magnon excitations in long-range ordered magnets. Compared to the neutron diffuse scattering that was used to characterize the quasi-elastic spiral spin-liquid correlations in the same compound~\cite{gao_spiral_2017}, the rich information that is available in inelastic neutron spectra allows a direct clarification of the further-neighbouring couplings in the spin Hamiltonian, which are crucial in understanding the phase transitions in MnSc$_2$S$_4$~\cite{bergman_order_2007, lee_theory_2008, iqbal_stability_2018}.

Figure~\ref{fig:ins_powder}b shows our inelastic neutron spectra collected on a powder sample of MnSc$_2$S$_4$ at temperature $T = 1.3 $~K in the helical ordered state, which is the parent phase of the field-induced triple-$\bm{q}$ state~\cite{gao_spiral_2017}. Strong inelastic scattering intensities are observed, emanating from the magnetic Bragg reflections that belong to the propagation vector $\bm{q}$ = (0.75 0.75 0), and reaching a maximal energy of $E \sim 0.9$ meV at wavevector $Q\sim 0.9$~\AA$^{-1}$.  Compared to other similar spinel compounds~\cite{zaharko_spin_2011, macdougall_revisiting_2016, ge_spin_2017}, the magnon dispersion bandwidth in MnSc$_2$S$_4$ is narrower, consistent with its relatively low ordering temperature of $T_N = 2.3$~K~\cite{fritsch_spin_2004,gao_spiral_2017}.

\begin{figure*}
\includegraphics[width=0.8\textwidth]{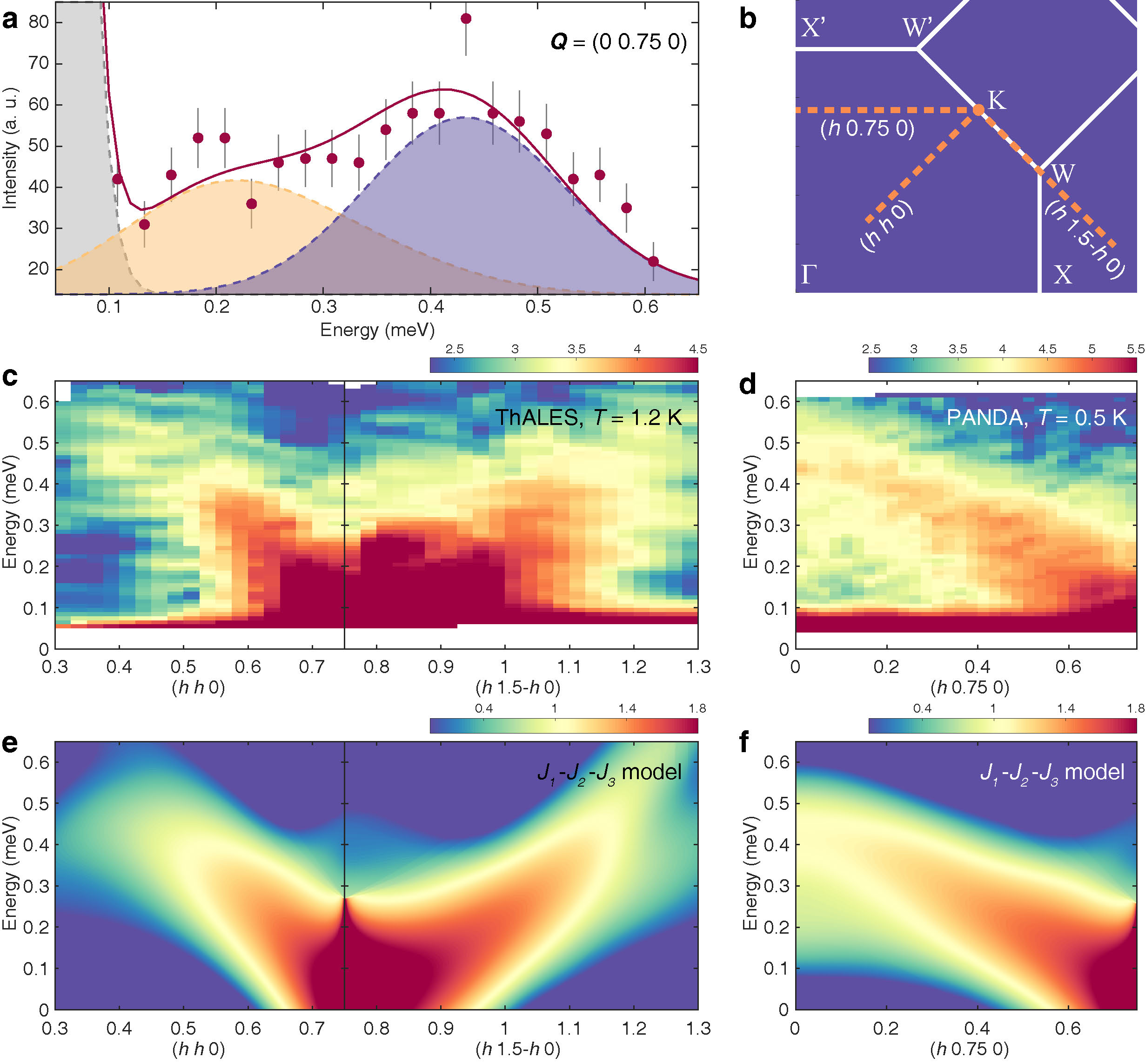}
\caption{\textbf{Spin dynamics in a single crystal sample of MnSc$_2$S$_4$. a,} Representative INS spectra $I(\omega)$ (red circles) collected on PANDA at $T= 0.5$ K and $\bm{Q}=$ (0 0.75 0). The red solid line denotes the calculated spectra using the $J_1$-$J_2$-$J_3$ model. Dashed lines with shaded areas indicate the contributions of magnon scattering from the (0.75 $\pm0.75$ 0) and (0 0.75 $\pm 0.75$) magnetic domains (yellow), magnon scattering from the (0.75 0 $\pm0.75$) magnetic domains (blue), and tail of the elastic line (grey). The calculated spectra are convoluted by a Gaussian function with fitted FWHM of 0.21 meV to account for the instrumental resolution and thermal broadening. \textbf{b,} Brillouin zone in the ($hk$0) plane with conventional notations. INS spectra are measured along the yellow dashed lines. \textbf{c,} INS spectra measured on ThALES at $T=1.2$ K along the ($h$ $h$ 0) and ($h$ 1.5$-h$ 0) directions. \textbf{d,} INS spectra measured on PANDA at $T=0.5$ K along the ($h$ 0.75 0) direction. \textbf{e,f,} Calculated INS spectra using the $J_1$-$J_2$-$J_3$ model. Error bars in \textbf{a} represent standard deviations.
\label{fig:ins_xtal}}
\end{figure*}

Figure~\ref{fig:ins_xtal} presents our INS results collected on a single crystal sample of MnSc$_2$S$_4$ along the high symmetry lines ($h$ $h$ 0), ($h$ $1.5\!-\!h$ 0), and ($h$ 0.75 0) in reciprocal space. No excitation gap can be resolved, which is compatible with the absence of single-ion anisotropy up to the second order in spin operators due to the $3d^5$ electron configuration of the Mn$^{2+}$ ions~\cite{watanabe_ground_1957}. A representative energy scan at (0 0.75 0) shown in Fig.~\ref{fig:ins_xtal}a reveals rather broad excitations, suggesting the appearance of multiple magnon bands.

Using linear spin wave theory, we are able to model the spin dynamics with Hamiltonian $\mathcal{H}_0 = \sum_{ij} J_{ij}\bm{S}_i \cdot \bm{S}_j$, where $J_{ij}$ is the exchange coupling between Heisenberg spins $\bm{S}_i$ and $\bm{S}_j$. As explained in Fig.~S1 of the Supplementary Information, it is necessary to include couplings up to the third-neighbours~\cite{bergman_order_2007,lee_theory_2008, iqbal_stability_2018} in order to reproduce the measured INS spectra. The fitted coupling strengths are $J_1 = -0.31(1)$ K, $J_2 = 0.46(1)$ K, and $J_3 = 0.087(4)$~K at the nearest-, second-,  and third-neighbours, respectively. Representative fits to the powder data at selected $Q$ positions are shown in Fig.~\ref{fig:ins_powder}d. The overall calculated spectra are presented in Fig.~\ref{fig:ins_powder}c and Fig.~\ref{fig:ins_xtal}e,f for comparison with the powder and single crystal experimental data, respectively. As shown in Fig.~\ref{fig:ins_xtal}a for the energy scan at (0 0.75 0), contributions from different magnetic domains are necessary to describe the broad excitations in the single crystal data.

Although the $J_1$-$J_2$-$J_3$ model successfully captures the spin dynamics in the helical phase of MnSc$_2$S$_4$, it fails to account for the field-induced triple-$\bm{q}$ phase~\cite{gao_spiral_2017}, which implies the necessity of even weaker perturbations that are beyond the INS resolution. Such a perturbation-dominated scenario is allowed in MnSc$_2$S$_4$ due to its enormous ground state degeneracy~\cite{bergman_order_2007, gao_spiral_2017}. Theoretical calculations on centrosymmetric systems have revealed that perturbations from the high-order analogs of the Ruderman-Kittel-Kasuya-Yosida (RKKY) interactions can often stabilize a triple-$\bm{q}$ phase~\cite{akagi_hidden_2012, hayami_effective_2017}. However, this mechanism fails in MnSc$_2$S$_4$ because the insulating character of this compound rules out any RKKY-like interactions that rely on the conduction electrons. 
\begin{figure}[t!]
\includegraphics[width=0.435\textwidth]{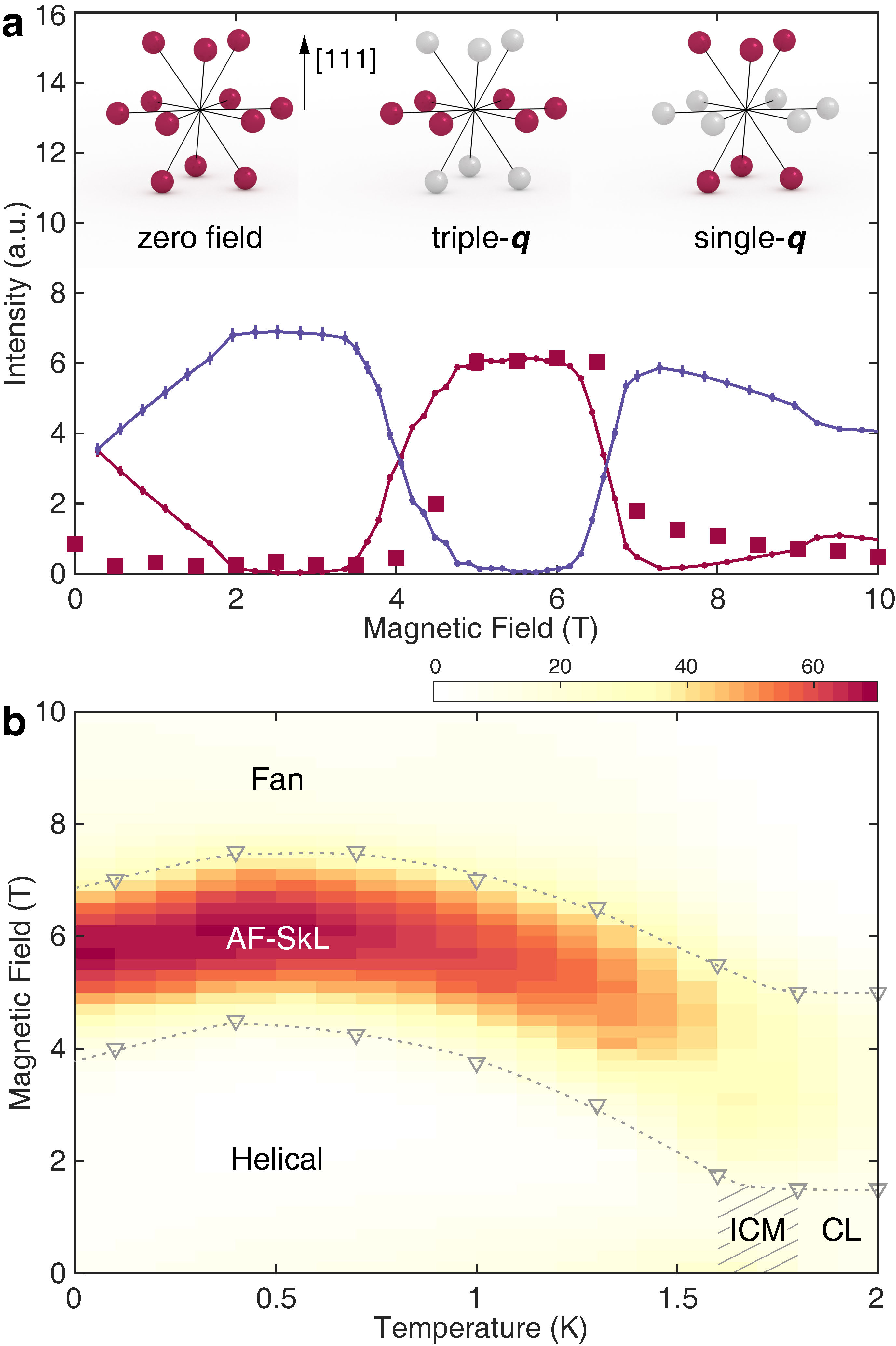}
\caption{\textbf{Anisotropic coupling induced triple-$\bm{q}$ phase in MnSc$_2$S$_4$. a,} Evolution of magnetic domains as a function of magnetic field applied along the [111] direction at $T= 0.1$ K. Red squares are neutron diffraction intensities of the (0.75 $-0.75$ 0) reflection within the (111) plane measured in a decreasing field. Solid lines are intensities obtained from Monte Carlo simulations, with the averaged contributions from the six arms perpendicular (non-perpendicular) to the [111] direction shown in red (blue). Error bars indicate the standard deviations of the mean. In the intermediate phase region between 3.5 and 7 T, the six arms perpendicular to the [111] direction have equal intensities, consistent with its triple-$\bm{q}$ character. Insets show the intensity distribution of the $\langle$0.75 0.75 0$\rangle$ star in the single-$\bm{q}$ helical phase observed in zero-field cooling (left), triple-$\bm{q}$ phase in an intermediate field (middle), and single-$\bm{q}$ helical phase with field-induced domain redistribution (right). Each dot represents a propagation vector, with red (grey) color indicating non-zero (zero) intensity.  \textbf{b,} Phase diagram for MnSc$_2$S$_4$ obtained from neutron diffraction experiment performed in a magnetic field along the [111] direction. Colormap shows the intensity of the (0.75 $-0.75$ 0) reflection collected in a decreasing field, and the phase boundary of the AF-Sk lattice state (AF-SkL) is marked by triangles that are connected by dashed lines as guide to the eyes. CL (ICM) stands for the single-$\bm{q}$ collinear (incommensurate) phase. 
The Fan phase is a single-$\bm{q}$ collinear phase added with a uniform magnetization along the [111] direction. Error bars representing the standard deviations are smaller than the marker size.
\label{fig:phase_diagram}}
\end{figure}

Through extensive Monte Carlo simulations, we explored the effect of different perturbations that are compatible with the symmetries of the lattice~\cite{lee_theory_2008}, and revealed that the triple-$\bm{q}$ phase in MnSc$_2$S$_4$ can be stabilized by anisotropic couplings at the nearest-neighbours together with a fourth-order single-ion anisotropy term that might be microscopically derived from the spin-orbit coupling and dipolar interactions~\cite{lee_theory_2008}. The perturbed $J_1$-$J_2$-$J_3$ Hamiltonian now reads

\begin{align}
\mathcal{H} &= \mathcal{H}_0 + \mathcal{H_{\parallel}} + \mathcal{H}_{\mathrm{A}} + \mathcal{H}_{\mathrm{Zeeman}}   \nonumber \\
 &= \sum_{ij} J_{ij}\bm{S}_i \cdot \bm{S}_j + 3J_{\parallel}\sum_{ij \in \mathrm{NN}}(\bm{S}_i\cdot \bm{\hat{r}}_{ij})(\bm{S}_j\cdot \bm{\hat{r}}_{ij}) \nonumber \\
 &+ A_4\sum_{i, \alpha = x,y,z} (S_i^\alpha)^4 - g\mu_B\bm{B}_{111}\sum_i \bm{S}_i\ \mathrm{,}
\end{align}
where  $\mathcal{H_{\parallel}}$ is the perturbation term due to the NN anisotropic couplings, in which $J_\parallel$ is the anisotropic coupling strength and $\bm{\hat{r}}_{ij}$ is the unitary direction vector along the NN bonds; $\mathcal{H}_\mathrm{A}$ describes a weak fourth-order single-ion anisotropy that is needed to stabilize a zero-field helical ground state~\cite{watanabe_ground_1957}; $\mathcal{H}_\mathrm{Zeeman}$ is the conventional Zeeman term for spins in a magnetic field $\bm{B}_{111}$ along the [111] direction. In our minimal Hamiltonian, the anisotropic $J_{\parallel}$ is found to be the only term that can induce a triple-$\bm{q}$ phase. Through comparison with the experimental phase diagram presented in Fig.~\ref{fig:phase_diagram}, the perturbation parameters can be determined to be $J_\parallel = -0.01$ K and $A_4 = 0.0016$ K. As exemplified in Fig.~\ref{fig:phase_diagram}a, only one triple-$\bm{q}$ domain with propagation vectors lying within the (111) plane is stabilized in field~\cite{gao_spiral_2017}, and the consequent non-monotonous evolution of the domain distribution is successfully reproduced in our simulations. As shown in Fig.~S3 of the Supplementary Information, the magnitude of the total scalar spin chirality increases sharply upon entering the triple-$\bm{q}$ phase, evidencing a magnetic structure that is topologically different from the single-$\bm{q}$ helical phase~\cite{okubo_multiple_2012, rosales_three_2015}.
\begin{figure*}[t!]
\includegraphics[width=0.95\textwidth]{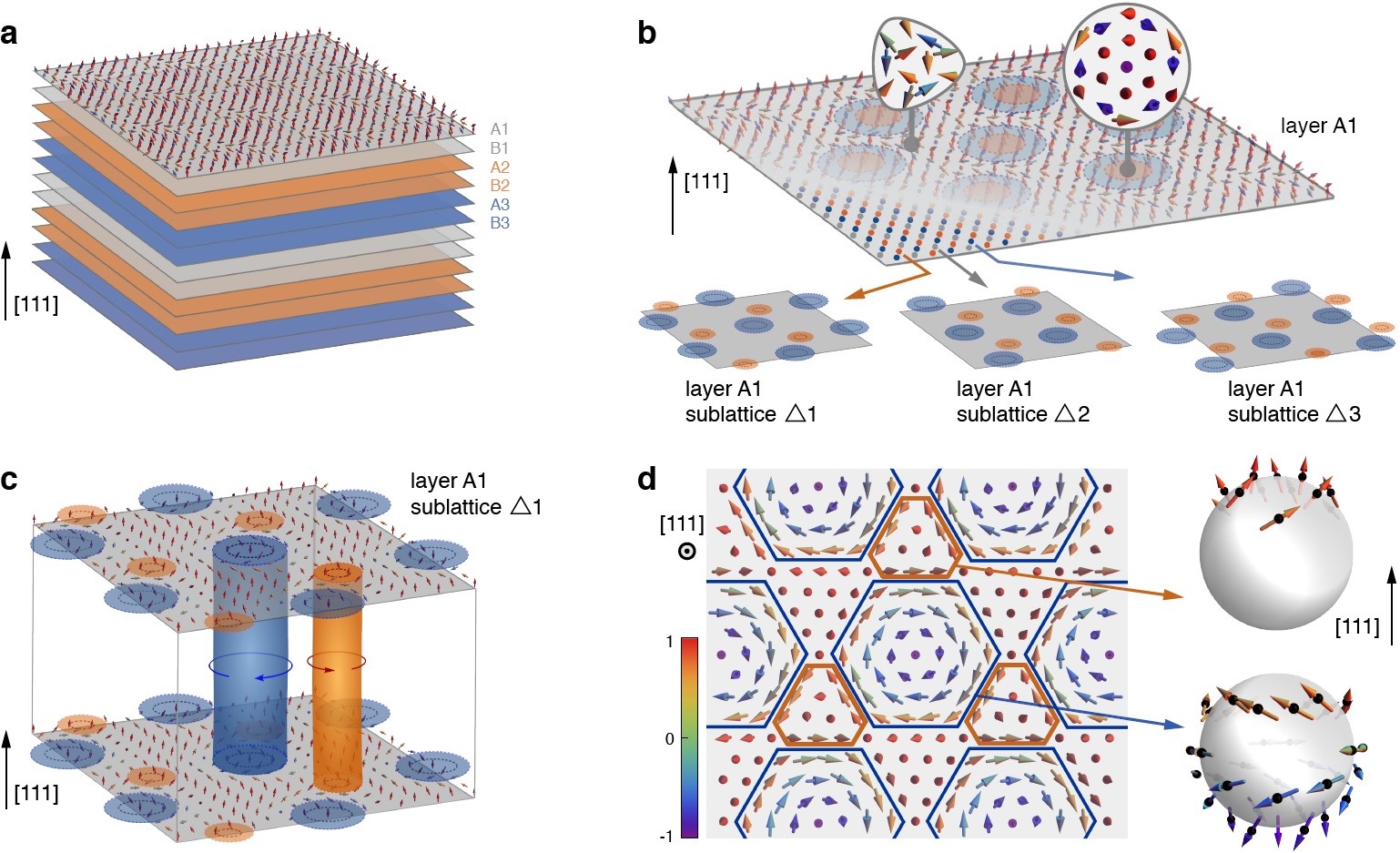}
\caption{\textbf{Fractional AF-Sk lattice in MnSc$_2$S$_4$. a,} Stacking order for the Mn$^{2+}$ triangular lattice layers along the [111] direction. A and B denote the two FCC sublattices of the diamond lattice shifted by (1/4 1/4 1/4), and the Mn$^{2+}$ positions are the same in the neighbouring layers of the same color. Within the A or B sublattice, the triangular lattices in the consecutive layers are shifted by (1/2 1/2 0), leading to three different types of layers shown by different colors. \textbf{b,} In each layer, the  Mn$^{2+}$ triangular lattice can be divided into three sublattices of $\triangle1$, $\triangle2$, and $\triangle3$ as illustrated in the insets.  In the triple-$\bm{q}$ phase, Mn$^{2+}$ spins within each triangular sublattice form a fractional Sk lattice (see panel \textbf{d}), which leads to a fractional AF-Sk lattice of three sublattices in the whole layer. Blue and yellow circles in the insets describe the locations of the fractional skyrmions with opposite winding direction, which overlap with each other in the complete (111) layer. The circular (triangular) inset is a zoomed-in plot of the magnetic structure around the skyrmion center (boundary). \textbf{c,} Spin configurations in the same type of layers are exactly the same, leading to cylinders of fractional skyrmions along the [111] direction. \textbf{d,} Spin texture of the triangular sublattice $\triangle1$ in layer A1. Directions of the spins are indicated by arrows, with colors denoting the size of the spin component along the [111] direction. Fractional skyrmions with opposite winding directions are indicated by blue and yellow lines in the main panel, and the wrapping of their spin texture are described over the two spheres shown on the right.
\label{fig:skyrmion}}
\end{figure*}

With Monte Carlo simulations, we can directly inspect the triple-$\bm{q}$ structure by layers, in which the Mn$^{2+}$ ions form a triangular lattice (see Figs.~\ref{fig:ins_powder}a and \ref{fig:skyrmion}a).  As expected for antiferromagnets, the spin configuration in one layer shown in Fig.~\ref{fig:skyrmion}b involve nearly anti-parallel spins at the nearest-neighbours. However, if the whole triangular lattice is separated into three sublattices~\cite{kamiya_magnetic_2014, rosales_three_2015} as shown in the insets of Fig.~\ref{fig:skyrmion}b, a smooth whirling texture will emerge in each sublattice, and the only difference among the sublattices is an overall shift of whorls. As described in Figs.~\ref{fig:skyrmion}c and d, spins at the centers of the whorls are anti-aligned with field, leading to a texture that is similar to the skyrmion lattices~\cite{muhlbauer_skyrmion_2009}. Due to the short distance between the centers of the whorls, skyrmions in the triangular sublattices are not wrapping the full sphere, but are fractionalized into two blocks with opposite winding directions~\cite{lin_skyrmion_2015}, forming a pair of incipient meron and antimeron~\cite{yu_transformation_2018} as indicated in Fig.~\ref{fig:skyrmion}d. When the three sublattices are added together as shown schematically in Fig.~\ref{fig:skyrmion}b, fractional skyrmions with opposite magnetizations overlap in the whole triangular lattice, leading to oscillating $S_{111}$ components near the center of the whorls and $120^\circ$-like alignments for the $S_{\perp}$ components close to the periphery, where $S_{111}$ ($S_{\perp}$) are magnetic moments along (perpendicular to) the [111] direction. Therefore, each (111) layer in the triple-$\bm{q}$ phase realizes a fractional AF-Sk lattice that is composed of three sublattices~\cite{rosales_three_2015}. 

Stacking of AF-Sk lattices along the [111] direction is determined by the propagation vectors and the Mn$^{2+}$ positions within the (111) layers. In the Methods section,  we present an analytical ansatz for spins at general positions constructed as a superposition of three helical modulations, and the correctness of the fractional AF-Sk lattice is verified through comparison against the neutron diffraction dataset shown in Fig.~S7 of the Supplementary Information.  The bipartite character of the diamond lattice leads to bilayers with exactly the same spin configurations as explained in Fig.~\ref{fig:skyrmion}a, thus realizing three consecutive AF-Sk bilayers with shifted whorl centers. Such a stacking order leads to AF-Sk tubes along the [111] direction shown in Fig.~\ref{fig:skyrmion}c, which is a common feature for many skyrmion lattices~\cite{milde_unwinding_2013,karube_robust_2016}.

The fractional AF-Sk lattice established in our work demonstrates that even antiferro-magnets can exhibit topologically non-trivial spin textures. In MnSc$_2$S$_4$, the AF-Sk lattice inherits the three-sublattice character of the triangular lattice in the (111) layers. However, the mechanism we discovered, which utilizes anisotropic couplings to stabilize a triple-$\bm{q}$ phase, can be generalized to AF systems with different geometries~\cite{attig_classical_2017, balla_affine_2019}. Especially, on the bipartite honeycomb lattice~\cite{gobel_anti_2017}, anisotropic couplings might stabilize a two-sublattice AF-Sk lattice with opposite spin winding textures, thus lending an ideal platform to explore the AF-Sk transport~\cite{barker_static_2016, zhang_antiferromagnetic_2016}.

The spin dynamics of the AF-Sks also deserves further investigations. In chiral systems, the lifetime of isolated AF-Sks is known to be enhanced by the DMI~\cite{bessarab_stability_2019}. It is therefore interesting to compare the effect of the antisymmetric couplings on the lifetime of the AF-Sks in centrosymmetric systems. For the AF-Sk lattice, magnons propagating through a topological spin texture might carry a Berry phase and thus experience a fictitious magnetic field~\cite{hoogdalem_magnetic_2013, daniels_topological_2019}, leading to the thermal Hall effect that can be utilized for magnonics applications. Furthermore, recent calculations on a three-sublattice AF-Sk lattice that is similar to the triple-$\bm{q}$ phase in MnSc$_2$S$_4$ revealed the lowest magnon band to be topological non-trivial~\cite{diaz_topological_2019}. The consequent chiral magnon edge states allow magnon transport without backscattering~\cite{roldan_topological_2016} and could further reduce the energy dissipation in magnonics devices.

In summary, our combined neutron scattering and Monte Carlo simulation works clarify the microscopic spin couplings in MnSc$_2$S$_4$ and establish the existence of a fractional AF-Sk lattice that is induced by the anisotropic couplings. Our work shows that topological structures can be stabilized in  antiferromagnets, which is an important step in fullfilling spintronic devices that aim to achieve efficient operations with a minimal scale. 

\section{Methods}

\textbf{Inelastic neutron scattering experiments.}
Inelastic neutron scattering experiments on a powder sample of MnSc$_2$S$_4$ were performed on FOCUS at the Swiss Spallation Neutron Source SINQ of the Paul Scherrer Institut PSI. For the measurements, about 4 g of MnSc$_2$S$_4$ powder sample synthesized through the solid-state reactions~\cite{krimmel_magnetic_2006} was filled into an annular-shaped aluminum can with outer/inner diameters of 12/10 mm. An orange cryostat with an additional roots pump was used, enabling a base temperature of 1.3 K. A setup with 5.0 \AA\ incoming neutron wavelength was employed.

Inelastic neutron scattering experiments on a single crystal sample of MnSc$_2$S$_4$ grown with the chemical transport reaction technique~\cite{gao_spiral_2017} were performed on ThALES~\cite{boehm_thales_2015,ill_data} at the Institut Laue-Langevin ILL and PANDA~\cite{schneidewind_panda_2015, utschick_optimizing_2016} at the Heinz Maier-Leibnitz Zentrum MLZ. Five crystals with a total mass of $\sim 100$ mg were co-aligned with $(hk0)$ as the horizontal scattering plane. For the experiment on ThALES, a cryomagnet together with an additional roots pump was used, which enabled a base temperature of 1.2 K and a maximal vertical field of 10~T. For better resolution, the Si(111) monochromator and PG(002) analyzer with double focusing were used. A Be-filter between the sample and analyzer and a radial collimator between the analyzer and detector were mounted. The final neutron momentum $k_f$ was fixed at 1.3 \AA$^{-1}$. For the experiment on PANDA, a $^{3}$He cryostat was used, which enabled a base temperature of $\sim 0.5$ K. PG(002) monochromator and analyzer with double focusing were employed. The final neutron momentum $k_f$ was fixed at 1.3 \AA$^{-1}$. A cooled Be filter was mounted before the sample to remove the higher-order neutrons.

Linear spin wave calculations and fits for the INS spectra were performed using the SpinW package~\cite{toth_linear_2015}. Input data for the fits are the three integrated intensities $I(\omega)$ shown in Fig.~\ref{fig:ins_powder}b. The spin Hamiltonian of the $J_1$-$J_2$-$J_3$ model has the helical ground state with a propagation vector $\bm{q} =$ (0.75 0.75 0).

\textbf{Neutron diffraction experiments.}
Neutron diffraction experiment was performed on the diffractometer D23 at the ILL to map out the phase diagram shown in Fig.~\ref{fig:phase_diagram}. Incoming neutron wavelength of 1.27 \AA\ was selected by the Cu(200) monochromator. A dilution refrigerator with a base temperature of 50 mK together with a magnet that supplies a field up to 12 T was employed. The MnSc$_2$S$_4$ crystal was aligned with the (111) direction along the vertical field direction. To map out the phase diagram, we first cooled the crystal in zero field, then perform rocking scan for the (0.75 $-0.75$ 0) reflection with increasing and decreasing fields.

The neutron diffraction dataset in the triple-$\bm{q}$ phase was collected on TriCS (now ZEBRA) at SINQ, PSI. Incoming neutron wavelength of 2.32~\AA\ was selected by the PG(002) monochromator. A PG filter was mounted before the sample. A cryomagnet together with a roots pump was employed for the measurements. 67 reflections were collected at $T$ = 1.60~K in a magnetic field of 3.5~T along the [111] direction.

\textbf{Monte Carlo simulations.}
Monte Carlo simulations were performed using the Metropolis algorithm by lowering the temperature in an annealing scheme and computing 500 independent runs initialized by different random numbers for each temperature and external magnetic field.  Simulations were performed in $2\times L^3$ magnetic site clusters, with $L=8-24$ and periodic boundary conditions. In order to compare the classical MC simulations with the experimental results, the $S^2$ factor in the computed thermal averages of relevant quantities was replaced by the quantum mechanical expectation value $\langle S^2\rangle = S(S + 1)$ following Ref.~\citenum{johnston_magnetic_2011}. 

\section{Acknowledgements}
We acknowledge S. T\'oth and S. Ward for help in the analysis of the neutron spectra. We thank A. Scaramucci for the initial trial on the Monte Carlo simulations. We acknowledge helpful discussions with M. Pregelj, S.B. Lee, T.-h. Arima, T. Nakajima, J.S. White, and Y. Su. S.G. acknowledges fruitful discussions at RIKEN CEMS. F.A.G.A. and H.D.R  thank R. Borzi for fruitful discussions. H.D.R. thanks M. Zhitomirsky for helpful discussions about the Monte Carlo simulations. Our neutron scattering experiments were performed at the Swiss Spallation Neutron Source SINQ, Paul Scherrer Institut, Villigen, Switzerland, the Heinz Maier-Leibnitz Zentrum MLZ, Garching, Germany, and the Institut Laue-Langevin ILL, Grenoble, France. 
This work was supported by the Swiss National Science Foundation under Grants No. 20021-140862, No. 20020-152734, the SCOPES project No. IZ73Z0-152734/1, and Centro Latinoamericano-Suizo under the Seed money Grant No. SMG1811. Our work was additionally supported by the Deutsche Forschungsgemeinschaft by the Transregional Collaborative Research Center TRR 80. D.C.C., F.A.G.A., and H.D.R. are partially supported by CONICET (PIP 2015-813), SECyT UNLP PI+D X792 and X788, PPID X039. H.D.R. acknowledges support from PICT 2016-4083.

\section{Author contributions}
O.Z. designed and coordinated the project. V.T. prepared the single crystals. S.G., O.Z., and C.R. performed the inelastic neutron scattering experiments with T.F. as the local contact for FOCUS, P.S. and M.B. for ThALES, P.C. and A.S. for PANDA. S.G. analyzed the neutron spectra with input from O.Z., T.F., and C.R. Neutron diffraction experiments were performed by G.K. and O.Z with E.R. as the local contact. Theoretical analysis and Monte Carlo simulations were performed by H.D.R., F.G.A., and D.C.C. The manuscript was prepared by S.G., H.D.R., and O.Z. with input from all co-authors.

\clearpage
\newpage

\renewcommand{\thefigure}{S\arabic{figure}}
\renewcommand{\thetable}{S\arabic{table}}

\renewcommand{\theequation}{\arabic{equation}}

\makeatletter
\renewcommand*{\citenumfont}[1]{S#1}
\renewcommand*{\bibnumfmt}[1]{[S#1]}
\def\clearfmfn{\let\@FMN@list\@empty}  
\makeatother
\clearfmfn

\setcounter{figure}{0} 
\setcounter{table}{0}
\setcounter{equation}{0} 

\onecolumngrid
\begin{center} {\bf \large Fractional antiferromagnetic skyrmion lattice induced by anisotropic couplings \\
 Supplementary Information} \end{center}
\vspace{0.5cm}

\maketitle

\textbf{Comparison for different spin models}

\begin{figure}[h!]
\includegraphics[width=0.98\textwidth]{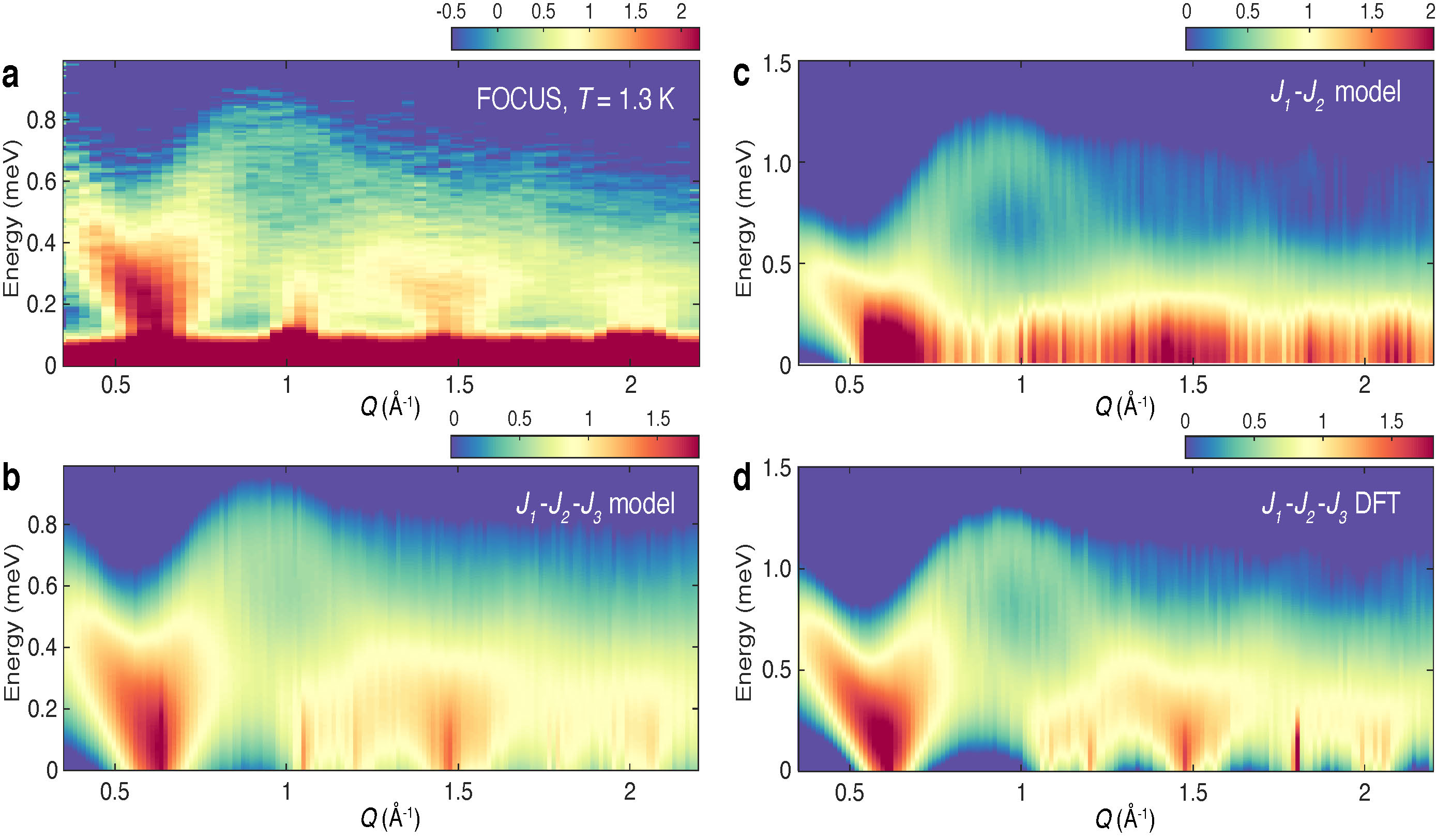}
\caption{\textbf{Comparison of different spin models. a,} INS spectra $S(Q,\omega)$ collected on FOCUS at $T=$ 1.3 K using a powder sample of MnSc$_2$S$_4$. \textbf{b-d} INS spectra calculated using the linear spin wave theory for the $J_1$-$J_2$-$J_3$ model with $J_1 = -0.31$ K, $J_2=0.46$ K, and $J_3=0.087$ K as presented in the main text (\textbf{b}), for the $J_1$-$J_2$ model with $J_1 = -0.71$ K, $J_2= -0.85\times J_1 = 0.60$ K (\textbf{c}), and for the $J_1$-$J_2$-$J_3$ model with parameters calculated from the DFT calculations~\cite{iqbal_stability_2018s} $J_1 = -0.378$ K, $J_2=0.621$ K, and $J_3=0.217$ K. Please note the different energy ranges in different panels.
\label{fig:ins_compare}}
\end{figure}

\begin{figure}[h!]
\includegraphics[width=0.35\textwidth]{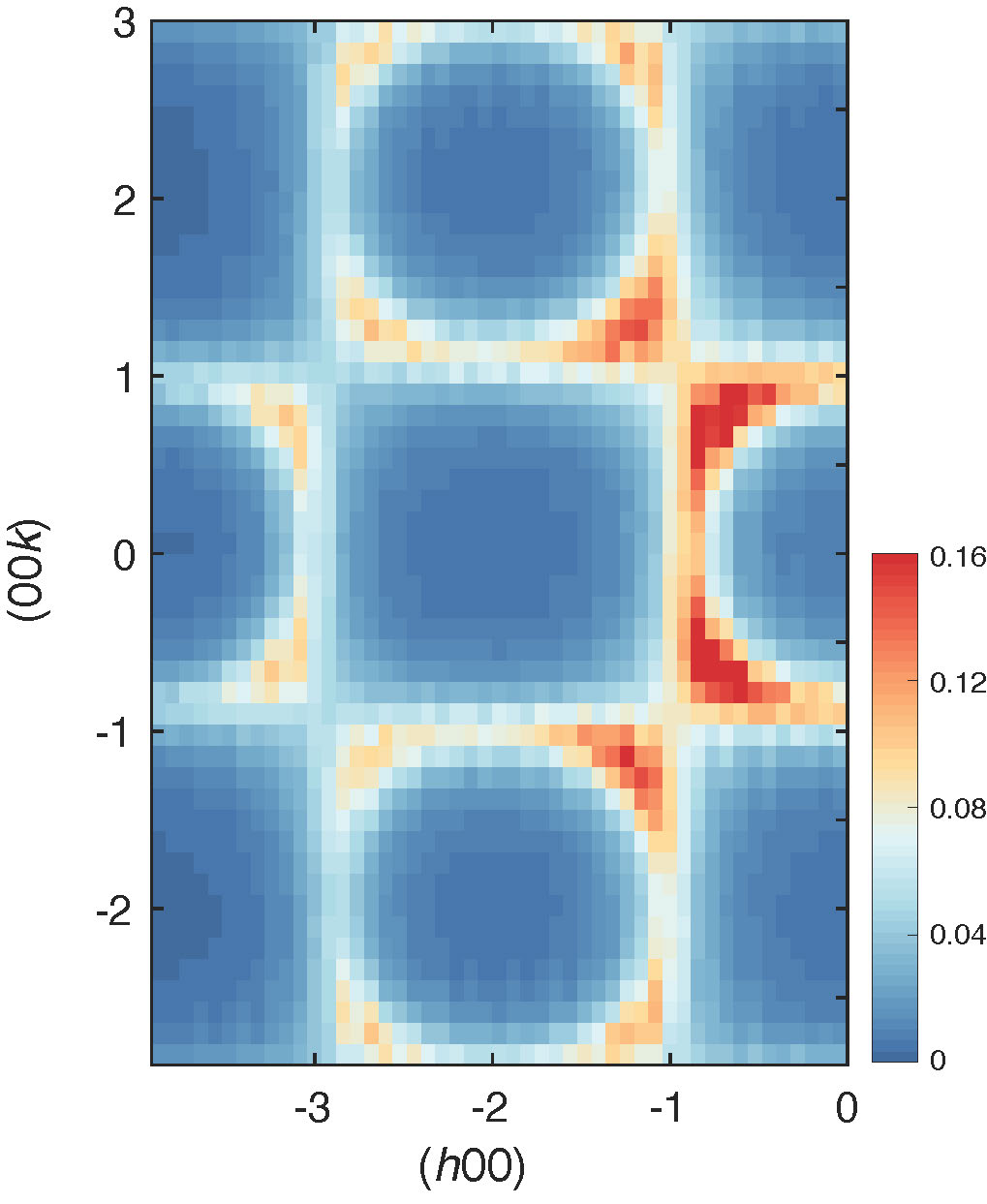}
\caption{\textbf{Spiral surface above the long-range order transition.} Spin correlations in the $(hk0)$ plane calculated by Monte Carlo simulations using the $J_1$-$J_2$-$J_3$ model plus the anisotropic perturbation terms with coupling strength listed in the main text. Calculations were performed at $T= 2.9$~K. Calculations with zero anisotropic perturbations does not affect the results.
\label{fig:diffuse}}
\end{figure}

Using linear spin wave theory, we compared different spin models against the INS spectra collected on a powder sample of MnSc$_2$S$_4$. Fig.~\ref{fig:ins_compare}a and b reproduce the experimental data and the spin wave calculation results for the $J_1$-$J_2$-$J_3$ model with $J_1 = -0.31$ K, $J_2=0.46$ K, and $J_3=0.087$ K as presented in the main text,  respectively. For the $J_1$-$J_2$ model with $J_3 = 0$, if the spectra at $Q\sim 0.4$~\AA\  was fitted to the experimental data, the calculated INS intensity will reach $\sim 1.2$ meV at $Q\sim 0.9$~\AA\ as shown in Fig.~\ref{fig:ins_compare}c, which is higher than the experimental bandwidth of $\sim 0.9$ meV. Therefore, the third-neighbour coupling $J_3$ is necessary to achieve a good fit for the INS spectra. The ratio $J_2/J_1$ is now increased to $\sim1.5$ as compared to 0.85 from neutron diffuse scattering~\cite{gao_spiral_2017s}, indicating that the lattice is even more frustrated than anticipated before. As shown in Fig.~\ref{fig:diffuse}, at temperatures above $T_N$, the  $J_1$-$J_2$-$J_3$ model leads to stronger intensities at around $\bm{q}=($0.75 0.75 0), which reproduces the intensity contrast within the spiral surface that was observed in our previous experiment~\cite{gao_spiral_2017s}.

Recent density functional theory (DFT) calculations~\cite{iqbal_stability_2018s} suggest a different $J_1$-$J_2$-$J_3$ model with $J_1 = -0.378$ K, $J_2=0.621$ K, and $J_3=0.217$ K. From the calculated INS spectra shown in Fig.~\ref{fig:ins_compare}d, we see that this DFT model produces a magnon bandwidth that is higher than the experimental observation. Compared to the coupling strengths fitted from the spin wave dispersions, the DFT model overestimates the coupling strength for $J_2$ and $J_3$.


\textbf{Theoretical phase diagram from the Monte Carlo simulations}

\begin{figure}[h!]
\includegraphics[width=0.5\textwidth]{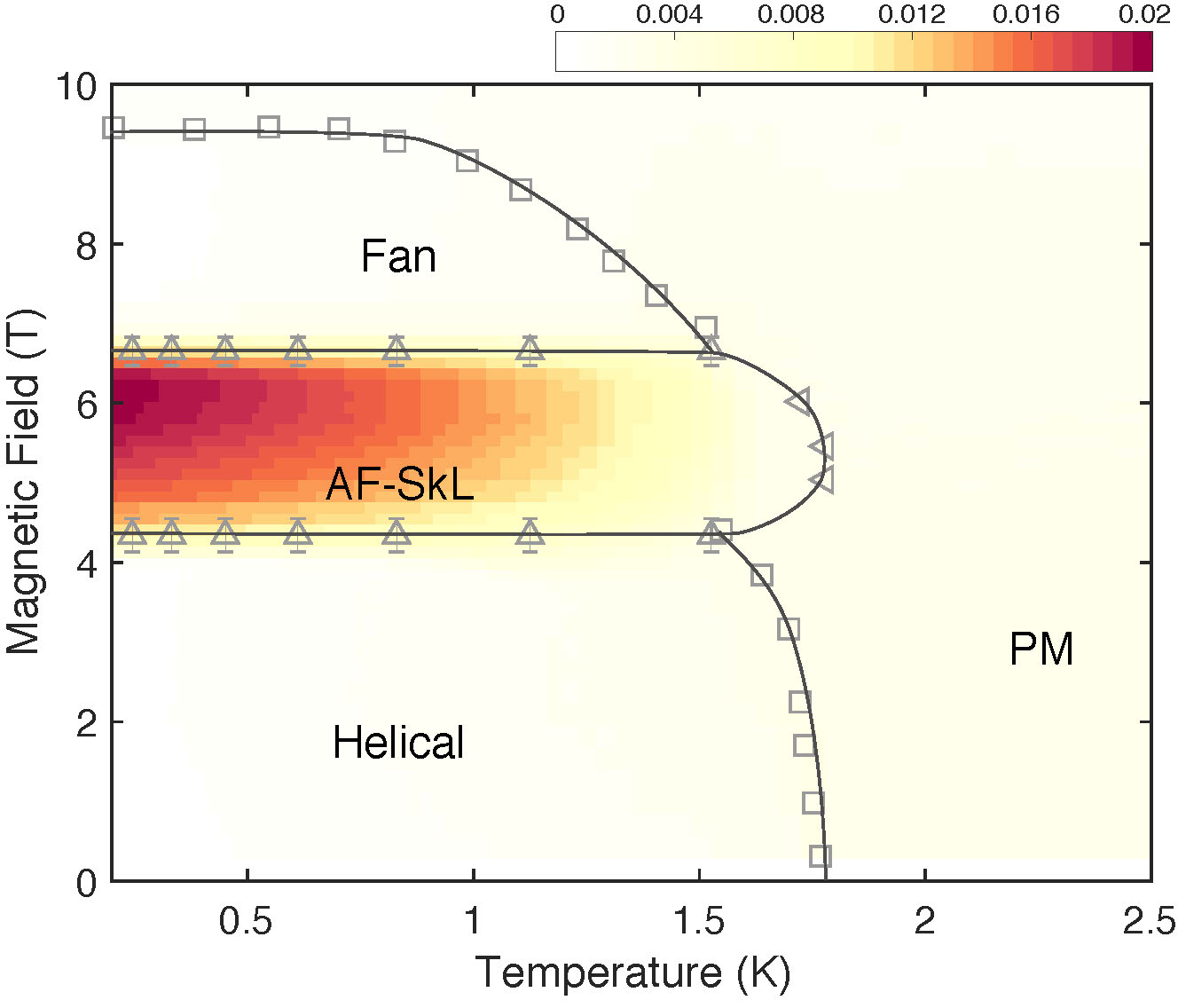}
\caption{\textbf{Calculated phase diagram with perturbations $J_{\parallel} = -0.01$~K and $A_4=0.0016$~K.} Phase diagram for MnSc$_2$S$_4$ obtained from the Monte Carlo simulation with field applied along the [111] direction as in the experiment. Colormap shows the calculated absolute value of the total scalar spin chirality $\chi_{\mathrm{tot}}$. Squares indicate the phase boundary obtained from the peak position of the calculated magnetic susceptibility in field along the [111] direction. Up-pointing triangles on the boundary of the AF-SkL phase are the middle points of the steep rise/drop in $\chi_{\rm{tot}}(H)$ at constant $T$, and their errors are estimated using the half-width of the transitional region.  Left-pointing triangles mark the sudden rise in $\chi_{\rm{tot}}(T)$ in constant field.  Error bars representing the standard deviations are not shown if their length is smaller than the marker size.
\label{fig:pd_calc}}
\end{figure}

Fig.~\ref{fig:pd_calc} plots the calculated phase diagram obtained from Monte Carlo simulations using the perturbed spin Hamiltonian (Eq. 1 in the main text). The $J_1$, $J_2$, and $J_3$ couplings are fixed to the spin wave fits of the INS spectra, while the anisotropy terms are determined to be $J_{\parallel} = -0.01$~K and $A_4=0.0016$~K after exploring the stability of the triple-$\bm{q}$ phase as discussed below. The color scale denotes the absolute value of the total scalar spin chirality $\chi_\mathrm{tot} = \langle \frac{1}{8\pi} \sum_n \chi_n \rangle$ with $\chi_n = \bm{S}_{i}\cdot (\bm{S}_{j} \times \bm{S}_{k})$, where $n$ indexes the $N$ elementary triangles of sites $i$, $j$, and $k$ in the (111) layers. 

In zero magnetic field the single-$\bm{q}$ helical state is identified by $\chi=0$. The transient collinear and incommensurate phases  found experimentally in  the vicinity of $T_N$ (Ref.~\cite{gao_spiral_2017s}) are not reproduced in our simulations possibly due to thermal fluctuations and finite size effects, and a detailed exploration in the transitional regime is deferred for future analysis.
In applied magnetic fields the triple-$\bm{q}$ phase is identified by sharp increase of the total scalar spin chirality, which evidences a magnetic structure that is topologically different from the single-$\bm{q}$ helical phase. Contrary to the skyrmion lattice that are stabilized by the antysimmetric Dzyaloshinskii-Moriya interactions~\cite{muhlbauer_skyrmion_2009s}, here the winding direction can be either clockwise or anti-clockwise since the model preserves the inversion symmetry in the (111) plane~\cite{okubo_multiple_2012s}. This implies a spontaneous symmetry breaking in the AF-SkL phase.

\begin{figure}
\includegraphics[width=0.7\textwidth]{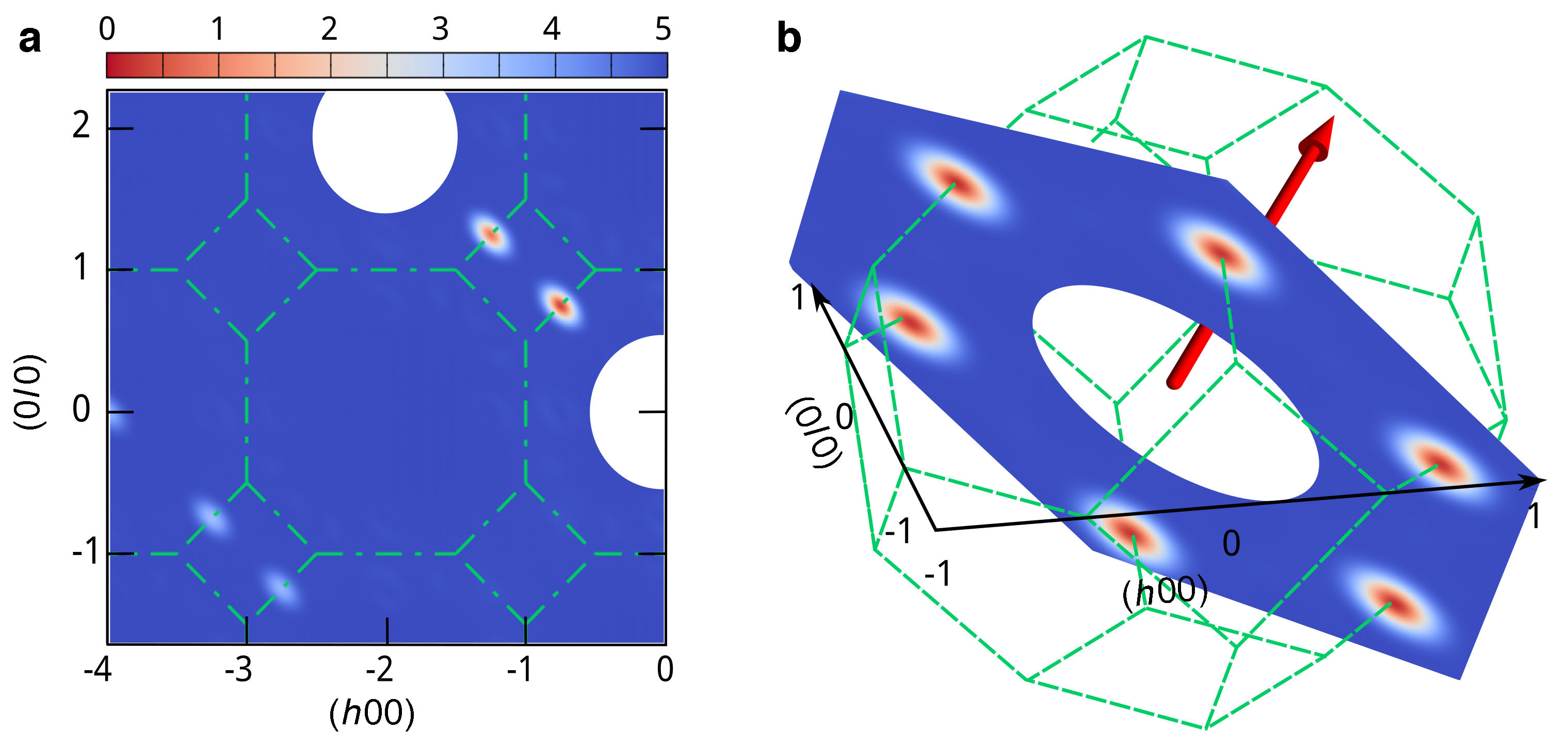}
\caption{\textbf{Identifying the triple-$\bm{q}$ phase.} Magnetic structure factor obtained by simulations in the triple-$\bm{q}$ phase at $T = 1.25$ K and $B_{111}=5.6$ T in the $(hk0)$ (\textbf{a}) and $(111)$ (\textbf{b}) planes. \label{fig:Sq_mc}}
\end{figure}

Two complementary methods have been employed to clarify the AF-SkL state in the Monte Carlo simulations. One is to directly check the magnetic textures in real space as exemplified in Fig.~4 of the main text, another is to calculate the magnetic structure factors in reciprocal space that can be directly compared to the neutron diffraction results. In the latter method, the skyrmion phase can be identified by the six Bragg spots located in the plane perpendicular to the magnetic field. Figure~\ref{fig:Sq_mc} shows the calculated magnetic structure factors in $(hk0)$ and $(111)$ planes at $T = 1.25$~K and $B_{111}=5.6$~T using a $16\times 16\times 16$ super-lattice over 500 averaged copies.

\begin{figure}[h!]
\includegraphics[width=0.85\textwidth]{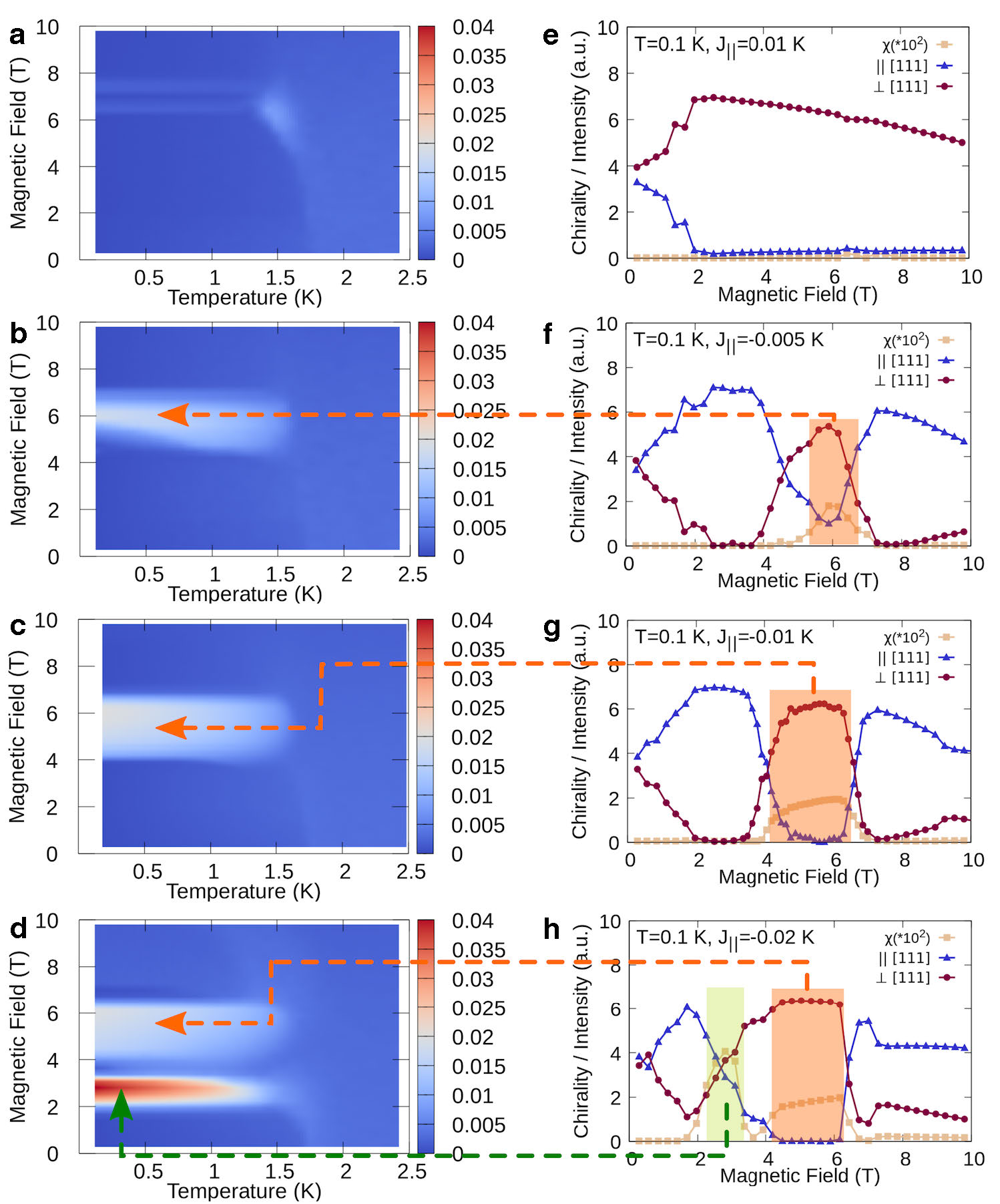}
\caption{\textbf{Dependence of the triple-$\bm{q}$ phase stability on the perturbation terms $J_{\parallel}$.} \textbf{a-d,} Calculated phase diagrams with perturbations $J_{\parallel} = 0.01$~K (\textbf{a}), $-0.005$~K (\textbf{b}), $-0.01$~K (\textbf{c}), and $-0.02$~K (\textbf{d}). The single-ion anisotropy $A_4$ is fixed at 0.0016 K. Colormap shows the absolute value of the total scalar spin chirality similar to that in Fig.~\ref{fig:pd_calc}. \textbf{e-h,}  Field dependence of the domain population at $T = 0.1$~K. Red circles (blue triangles) indicate domains with $\bm{q}$ in (out of) the (111) plane. Yellow squares are the calculated absolute value of the scalar spin chirality. Error bars representing the standard deviations of the mean are smaller than the marker size.
\label{fig:MC_expand}}
\end{figure}

In order to illustrate the stability of the triple-$\bm{q}$ phase and explain how did we determine the strength of the perturbation terms, we compare the phase diagrams calculated with different strength of $J_{\parallel}$ in Fig.~\ref{fig:MC_expand}. When the strength of $J_{\parallel}$ is reduced from $-0.01$~K to $-0.005$~K, the stability region of the triple-$\bm{q}$ phase will also become reduced and thus deviates from our experimental observation. On the other hand, when the strength of $J_{\parallel}$ is increased to $-0.02$~K, although  the stability region of the triple-$\bm{q}$ phase remains almost the same, a new chiral phase emerges at lower magnetic fields, which is possibly a multiple-$\bm{q}$ state that is different from the skyrmion, fractional skyrmion, or meron lattices. Finally, when the sign of $J_\parallel$ become positive with $J_{\parallel}=0.01$~K, the triple-$\bm{q}$ phase will disappear completely. Therefore, the perturbation term $J_{\parallel}$ can be determined to be $-0.01$~K.

\bigbreak
\textbf{Analytical expression for the AF-Sk lattice}

As confirmed in many different types of skymion lattices, the magnetic structure of each $\bm{q}$-component of the triple-$\bm{q}$ structure is often related to the single-$\bm{q}$ structure observed in zero field. A well-known example is the Bloch-type skyrmion lattice in MnSi (Ref.~\citenum{muhlbauer_skyrmion_2009s}), where the helical components are derived from the zero-field helical phase. Similar arguments hold for the cycloidal components of the N\'eel-type skyrmion lattice observed in GaV$_4$S$_8$ (Ref.~\citenum{kezsmarki_neel_2015}). Therefore, considering the helical and collinear structures that are observed in MnSc$_2$S$_4$ at zero field, we can express its field-induced triple-$\bm{q}$ structure through the ansatz:
\begin{align}
\bm{S}(\bm{r}) &= \frac{1}{n_S}(A_\perp\sum_{i=1}^3 \sin(\bm{q}_i\cdot \bm{r}+\phi_{\perp})\bm{\hat{e}}_i \nonumber \\
&+ A_{111}\sum_{i = 1}^3 \cos(\bm{q}_i\cdot \bm{r} + \phi_{111}) \bm{\hat{e}}_{111} \nonumber \\      
&+ \bm{M}_{111})\ \mathrm{,}
\end{align}
where $n_S$ is the normalization factor that fixes the spin magnitude to $5/2$, $A_\perp$ ($A_{111}$) is the amplitude for spin modulation perpendicular (parallel) to the [111] direction $\bm{\hat{e}}_{111}$ with phase factor $\phi_{\perp,i}=\phi_\perp$ ($\phi_{111,i}=\phi_{111}$), $\bm{q}_i$ are the three propagation vectors (0.75 $-0.75$ 0), (0.75 0 $-0.75$), and (0 0.75 $-0.75$), $\bm{\hat{e}}_i$ are the unitary vectors that form cartesian coordinate systems with the corresponding $\bm{q}_i$ and $\bm{\hat{e}}_{111}$, and $\bm{M}_{111}$ is an homogeneous contribution to the magnetization along $\bm{\hat{e}}_{111}$. 

\begin{figure}[t!]
\includegraphics[width=0.95\textwidth]{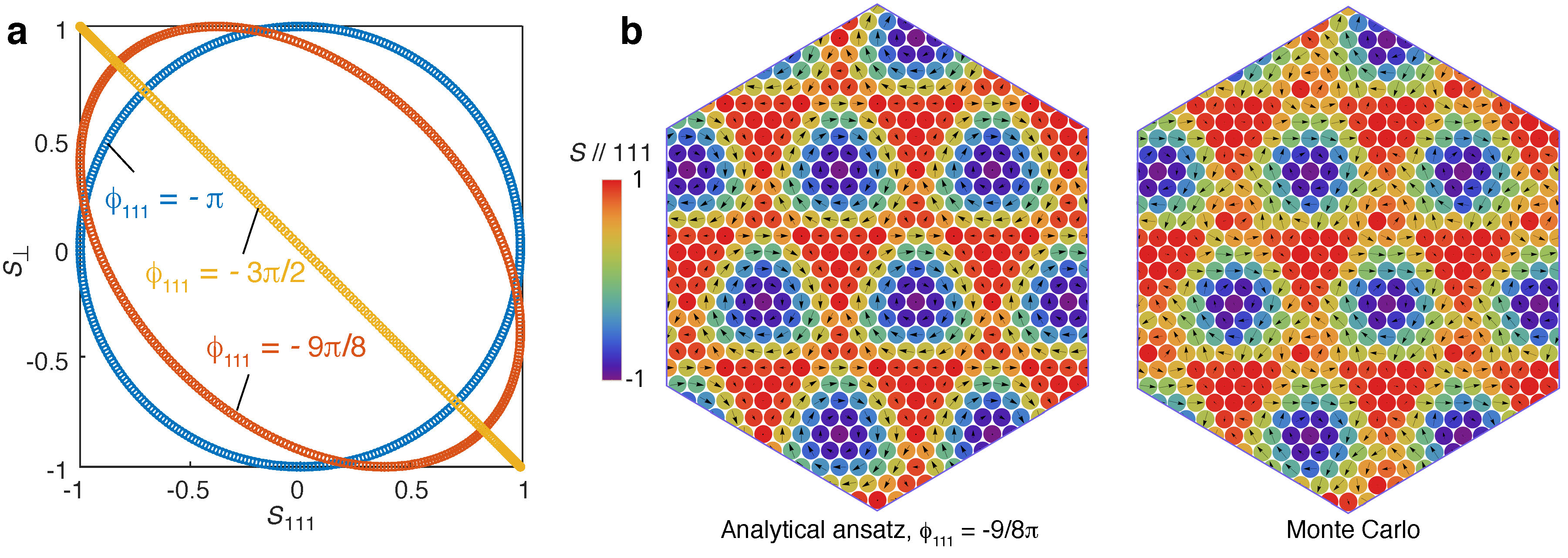}
\caption{\textbf{Analytical ansatz for the AF-Sk lattice.} \textbf{a,} Schematic for the moment directions in each $\bm{q}$-component of the triple-$\bm{q}$ structure at $\phi_{111}=-\pi$  (helical), $-3/2\pi$ (collinear), and $-9/8\pi$ (distorted helical). \textbf{b,} Comparison between representative magnetic texture for one sublattice in the (111) plane obtained by  the analytical ansatz (left) and the Monte Carlo simulations (right) performed at $T = 0.5$~K and $B_{111}=5$~T. The color scheme indicates the spin component along the [111] direction, and the arrows indicate the spin component in the (111) plane. 
\label{fig:ansatz}}
\end{figure}

Assuming equal magnitude for $A_\perp$ and $A_{111}$, and $\phi_{\perp} = 0$ without loss of generality, the case of $\phi_{111} = -\pi$ and $-3 \pi /2$ corresponds to helical and collinear components, respectively (see Fig.~\ref{fig:ansatz}a). Note that for the zero-field collinear structure, the spin directions are canted out of the (111) plane by $45^\circ$ according to our previous refinement~\cite{gao_spiral_2017s}, and such a canting has been taken into account in our expression. Therefore, by varying $\phi_{111}$, we can construct different triple-$\bm{q}$ structures with $\bm{q}$-components covering the observed collinear structure, helical structure, and most importantly, a general distorted structure that lies in-between the collinear and helical phases.

\begin{figure}[h!]
\includegraphics[width=0.82\textwidth]{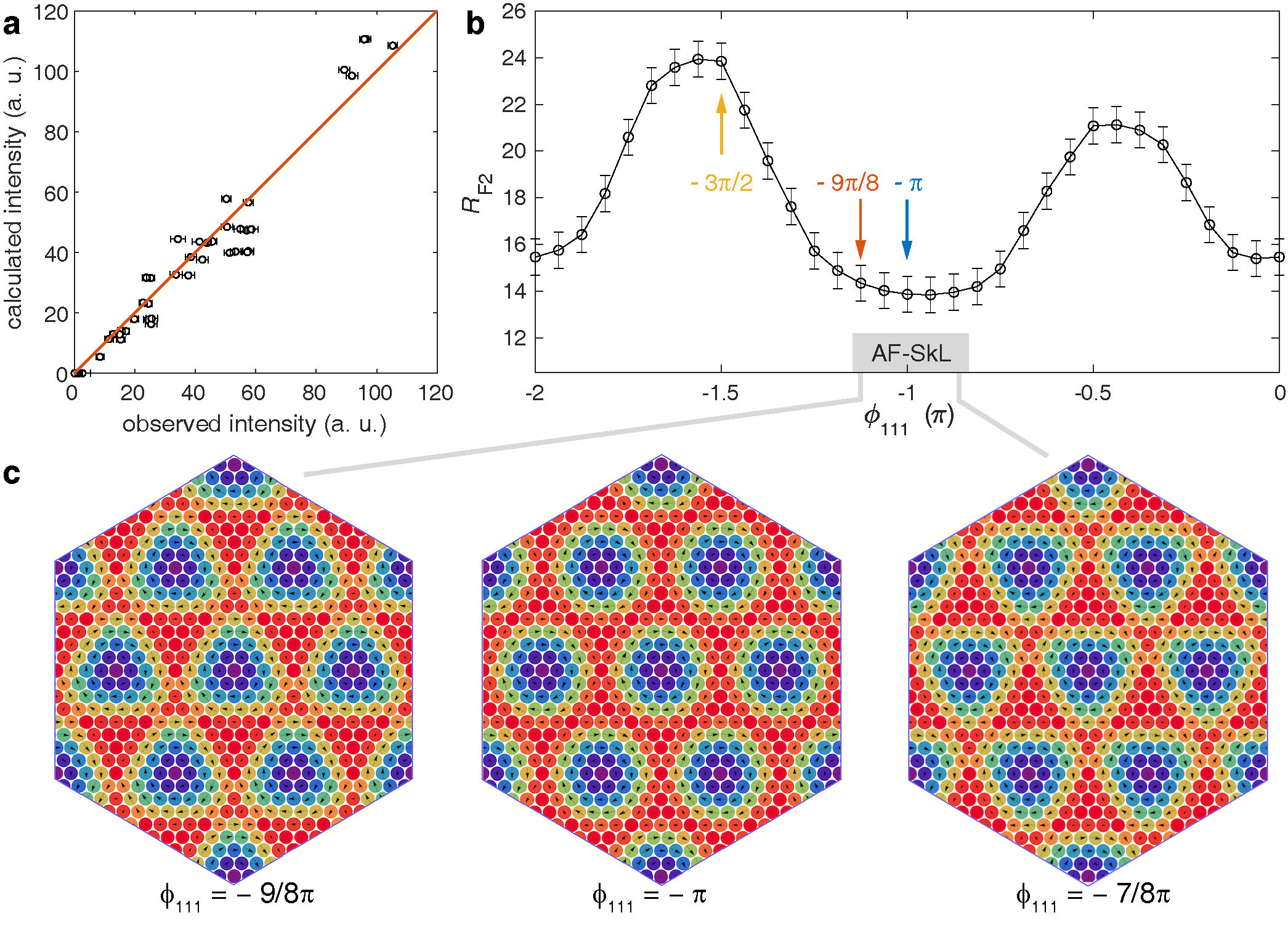}
\caption{\textbf{Refinement of the neutron diffraction dataset collected in triple-$\bm{q}$ phase.} \textbf{a,} Comparison of the observed and calculated intensities for the fractional antiferromagnetic skyrmion lattice. The dataset was collected in the triple-$\bm{q}$ phase under a magnetic field of 3.5 T along the [111] direction. \textbf{b,} Dependence of the $R_{F2}$ factor on the phase factor $\phi_{111}$. The arrows indicate results for $\phi_{111}= -\pi$, $-3/2\pi$, and $-9/8\pi$,  which correspond to the triple-$\bm{q}$ structures with helical, collinear, and distorted helical components, respectively. \textbf{c,} Magnetic textures for one sublattice in the (111) plane with $\phi_{111}= -9/8\pi$, $-\pi$, and $-7/8\pi$, showing that in the region of $-9/8\pi \leq \phi_{111} \leq -7/8\pi$, the triple-$\bm{q}$ structure always realizes a fractional AF-SkL and only the proportion of fractionalization is varied.  The color scheme indicates the spin component along the [111] direction, and the arrows indicate the spin component in the (111) plane.
\label{fig:refine}}
\end{figure}

Figure \ref{fig:ansatz}b shows the representative magnetic structure of the proposed ansatz for one sublattice in the (111) plane together with that obtained from the Monte Carlo simulations. Assuming $|\bm{M}_{111}|=1$, the parameter set of $A_{111} = -A_\perp = 2.2$, $\phi_\perp = 0$, and $\phi_{111} = -9\pi/8$, the proposed ansatz well reproduces the magnetic structure obtained in the Monte Carlo simulations. Two very important details can be observed from this result. First, unlike what happens in the typical skyrmion lattice, the internal phase for the spin configuration is different for the perpendicaular and parallel component of the spin $\phi_{\perp,i}\neq \phi_{111,i}$. Secondly, the condition $\sum_i\cos\phi_i= 1$  is not satisfied as usual in triple-${\bm q}$ phases \cite{okubo_multiple_2012s}. 

\bigbreak
\textbf{Refinement of the neutron diffraction dataset in the triple-$\bm{q}$ phase}

With the ansatz presented in the previous section, we can directly verify the antiferro-magnetic skyrmion lattice by comparing its magnetic structure factors with the neutron diffraction intensities of magnetic Bragg peaks.  Details for the neutron diffraction experiment can be found in the Methods section. As shown in Fig.~\ref{fig:refine}a, the fractional antiferromagnetic skyrmion lattice obtained in the Monte Carlo simulation fits the neutron diffraction dataset very well, with $R$-factors $R_{F2} = 14.3~\%$ and $R_F = 10.8~\%$. 

By varying the $\phi_{111}$ phase factors, we compared the refinement results from different triple-$\bm{q}$ structures that are composed of general distorted helical components. Fig.~\ref{fig:refine}b summarizes the dependence of the $R_{F2}$ factor on $\phi_{111}$. The best refinement was achieved in the region of $-9/8\pi \leq \phi_{111} \leq -7/8\pi$ with comparable $R$-factors, justifying the value of $\phi_{111}=-9/8\pi$ obtained from the Monte Carlo simulations. More importantly, as shown in Fig.~\ref{fig:refine}c, in the whole regime of $-9/8\pi \leq \phi_{111} \leq -7/8\pi$, the triple-$\bm{q}$ structure can always be described as a fractional AF-SkL, that is, each (111) plane exhibit a three-sublattice antiferromagnetic alignment, and a fractional skyrmion lattice emerges in each sublattice. The only difference in these structures is a slight variation in the fractionalization. Therefore, our neutron diffraction results strongly support the emergence of a fractional three-sublattice AF-SkL in MnSc$_2$S$_4$.

%

\end{document}